\documentclass[12pt]{article}
\usepackage[T1]{fontenc} \usepackage[latin1]{inputenc}
\usepackage{amsmath} \usepackage{graphicx}
\usepackage{amssymb}

\begin{document}

\begin{center}
\begin{Large}
{\bf Anomalous Dynamics of Unbiased Polymer Translocation through a Narrow
Pore}
\end{Large}

\vspace{1cm}

Debabrata Panja$^{*}$, Gerard T. Barkema$^{\dagger,\ddagger}$ and
Robin C. Ball$^{**}$

\vspace{10mm}
$^*$Institute for Theoretical Physics, Universiteit van Amsterdam,\\
Valckenierstraat 65, 1018 XE Amsterdam, The Netherlands

$^{\dagger}$Institute for
Theoretical Physics, Universiteit Utrecht, Leuvenlaan 4,\\ 3584 CE
Utrecht The Netherlands

$^{\ddagger}$Instituut-Lorentz,
Universiteit Leiden, Niels Bohrweg 2,\\ 2333 CA Leiden, The
Netherlands

 $^{**}$Department of Physics, University of Warwick, Coventry CV4
7AL, UK

\vspace{10mm}

{\bf Abstract}

\vspace{3mm} 
\begin{minipage}{0.75\linewidth}
We consider a polymer of length $N$ translocating through a narrow
pore in the absence of external fields. Characterization of its
purportedly anomalous dynamics has so far remained incomplete. We show
that the polymer dynamics is anomalous until the Rouse time
$\tau_{R}\sim N^{1+2\nu}$, with a mean square displacement through the
pore consistent with $t^{(1+\nu)/(1+2\nu)}$, with $\nu\approx0.588$
the Flory exponent. This is shown to be directly related to a decay in
time of the excess monomer density near the pore as
$t^{-(1+\nu)/(1+2\nu)}\exp(-t/\tau_{R})$. Beyond the Rouse time
translocation becomes diffusive. In consequence of this, the dwell-time
$\tau_{d}$, the time a translocating polymer typically spends within
the pore, scales as $N^{2+\nu}$, in contrast to previous claims.

\vspace{5mm}
\end{minipage}
\end{center}

\vspace{5mm}
Transport of molecules across cell membranes is an essential mechanism
for life processes. These molecules are often long and flexible, and
the pores in the membranes are too narrow to allow them to pass through
as a single unit. In such circumstances, the passage of a molecule
through the pore --- i.e.\ its translocation --- proceeds through
a random process in which polymer segments sequentially move through
the pore. DNA, RNA and proteins are naturally occurring long molecules
\cite{drei} subject to translocation in a variety of biological processes.
Translocation is used in gene therapy \cite{szabo}, in delivery of
drug molecules to their activation sites \cite{tseng}, and as an
efficient means of single molecule sequencing of DNA and RNA \cite{nakane}.
Understandably, the process of translocation has been an active topic
of current research: both because it is an essential ingredient in
many biological processes and for its relevance in practical applications.

Translocation is a complicated process in living organisms --- its
dynamics may be strongly influenced by factors like the presence of
chaperon molecules, pH values, chemical potential gradients,
assisting molecular motors etc. \cite{wickner}. In studies of
translocation as a \emph{biophysical} process, the polymer is
simplified to a sequentially connected string of $N$ monomers. Herein,
the quantities of interest are the typical time for the polymer to
leave a confining cell or vesicle, the ``escape time''
\cite{sungpark1}, and the typical time the polymer spends in the pore
or ``dwell time'' \cite{sungpark2}, as a function of chain length $N$
and other parameters like membrane thickness, membrane adsorption,
electrochemical potential gradient, etc. \cite{lub}.  These have been
measured directly in numerous experiments \cite{expts}.
\begin{figure}[h]
\begin{center}
\includegraphics[width=0.5\linewidth]{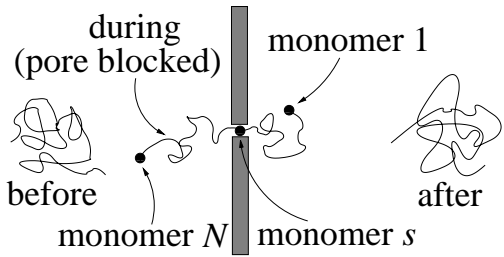}
\caption{Pictorial representation of a translocation event, with the
polymer shown before, during and after translocation.  We number the
monomers, starting with the end-monomer on the side it moves to. The
number of the monomer located in the middle of the pore is
$s$.\label{fig:sketch}}
\end{center}
\end{figure}

Experimentally, the most studied quantity is the dwell time
$\tau_{d}$, i.e., the pore blockade time for a translocation event
(see Fig. \ref{fig:sketch}). For theoretical descriptions of
$\tau_{d}$, during the last decade a number of mean-field type
theories \cite{sungpark1,sungpark2,lub} have been proposed, in which
translocation is described by a Fokker-Planck equation for
first-passage over an entropic barrier in terms of a single ``reaction
coordinate'' $s$. Here $s$ is the number of the monomer threaded at
the pore ($s=1,\ldots,N$). These theories apply under the assumption
that translocation is slower than the equilibration time-scale of the
entire polymer, which is likely for high pore friction. In Ref.
\cite{kantor}, this assumption was questioned, and the authors found
that for a self-avoiding polymer performing Rouse dynamics,
$\tau_{d}\ge\tau_{R}$, the Rouse time. Using simulation data in 2D,
they suggested that the inequality may actually be an equality, i.e.,
$\tau_{d}\sim\tau_{R}\sim N^{1+2\nu}$, which is $N^{2.5}$ in two
dimensions. Numerical data in support of this suggestion in 2D 
appeared in Ref. \cite{luo}. However, in a publication due to 
two of us, $\tau_{d}$ in 3D was numerically found to scale as $\sim
N^{2.40\pm0.05}$ \cite{wolt}, significantly larger than $N^{1+2\nu}$,
which is $N^{2.18}$ in three dimensions.  Additionally, in a recent
publication \cite{dubbeldam} $\tau_{d}$ was numerically found to scale
as $N^{2.52\pm0.04}$ in three dimensions (a discussion on the theory
of Ref. \cite{dubbeldam} appears at the end of this paper). Note that
these simulations do not incorporate hydrodynamical interactions, which 
are certainly important in experiments. Also, these simulations (and all 
theoretical studies, including this one) ignore interactions with other
polymers, i.e., they consider polymers in infinitely dilute solutions,
while in cell environments, the solution is not infinitely dilute. In
this paper we consider translocation in the absence of hydrodynamical
interactions, and at the end we reflect on the results we expect when
the hydrodynamical interactions are included. We also note here that
simulations with hydrodynamical interactions are non-trivial and costly.

Amid all the above results on $\tau_{d}$ mutually differing by
$\sim~O(N^{0.2})$, the only consensus that survives is that
$\tau_{d}\ge\tau_{R}$ \cite{kantor,wolt}.  Simulation results alone
cannot determine the scaling of $\tau_{d}$: different groups use
different polymer models with widely different criteria for
convergence for scaling results, and as a consequence, settling
differences of $\sim~O(N^{0.2})$ in $O(\tau_{R})$, is extremely
delicate.

An alternative approach that can potentially settle the issue of
$\tau_{d}$ scaling with $N$ is to analyze the dynamics of
translocation at a microscopic level. Indeed, the lower limit
$\tau_{R}$ for $\tau_{d}$ implies that the dynamics of translocation
is anomalous \cite{kantor}.  We know of only two published studies on
the anomalous dynamics of translocation, both using a fractional
Fokker-Planck equation (FFPE) \cite{klafter,dubbeldam}. However,
whether the assumptions underlying a FFPE apply for polymer
translocation are not clear. Additionally, none of the studies used
FFPE for the purpose of determining the scaling of $\tau_{d}$. In view
of the above, such a potential clearly has not been thoroughly
exploited.

The purpose of this paper is to report the characteristics of the
anomalous dynamics of translocation, \textit{derived from the
microscopic dynamics of the polymer}, and the scaling of $\tau_{d}$
obtained therefrom. Translocation proceeds via the exchange of
monomers through the pore: imagine a situation when a monomer from the
left of the membrane translocates to the right. This process increases
the monomer density in the right neighbourhood of the pore, and
simultaneously reduces the monomer density in the left neighbourhood
of the pore.  The local enhancement in the monomer density on the
right of the pore \textit{takes a finite time to dissipate away from
the membrane along the backbone of the polymer\/} (similarly for
replenishing monomer density on the left neighbourhood of the
pore). The imbalance in the monomer densities between the two local
neighbourhoods of the pore during this time implies that there is an
enhanced chance of the translocated monomer to return to the left of
the membrane, thereby giving rise to \textit{memory effects\/}, and
consequently, rendering the translocation dynamics subdiffusive. More
quantitatively, the excess monomer density (or the lack of it) in the
vicinity of the pore manifests itself in reduced (or increased) chain
tension around the pore, creating an imbalance of chain tension across
the pore (we note here that the chain tension at the pore acts as
monomeric chemical potential, and from now on we use both terms
interchangeably). We use well-known laws of polymer physics to show
that in time the imbalance in the chain tension across the pore
relaxes as $t^{-(1+\nu)/(1+2\nu)}\exp(-t/\tau_{R})$.  (Strictly
speaking, $\tau_{R}$ in this expression should be replaced by the
characteristic equilibration time of a tethered polymer with length of
$O(N)$; since both scale as $N^{1+2\nu}$, we use $\tau_{R}$ here,
favouring notational simplicity).  This results in translocation
dynamics being subdiffusive for $t<\tau_{R}$, with the mean-square
displacement $\langle\Delta s^{2}(t)\rangle$ of the reaction
coordinate $s(t)$ increasing as $t^{(1+\nu)/(1+2\nu)}$; and diffusive
for $t>\tau_{R}$. With $\sqrt{\langle\Delta
s^{2}(\tau_{d})\rangle}\sim N$, this leads to $\tau_{d}\sim N^{2+\nu}$.

We substantiate our theoretical derivations with extensive Monte Carlo 
simulations, in which the polymer performs single-monomer moves. The
definition of time is such that single-monomer moves along the
polymer's contour are attempted at a fixed rate of unity, while moves
that change the polymer's contour are attempted ten times less often.
Details of our self-avoiding polymer model in 3D can be found in
Refs. \cite{heukelum03,longpaper}.

The key step in quantitatively formulating the anomalous dynamics of
translocation is the following observation: a translocating polymer
comprises of \textit{two polymer segments tethered at opposite ends of
the pore\/} that are able to exchange monomers between them through
the pore; so \textit{each acts as a reservoir of monomers for the
other.} The velocity of  translocation $v(t)=\dot{s}(t)$, representing
monomer current, responds to $\phi(t)$, the imbalance in the monomeric
chemical potential across the pore acting as
``voltage''. Simultaneously, $\phi(t)$ also adjusts in response to
$v(t)$. In the presence of memory effects, they are related to each
other by $\phi(t)=\int_{0}^{t}dt'\mu(t-t')v(t')$ via the memory kernel
$\mu(t)$, which can be thought of as the (time-dependent) `impedance'
of the system. Supposing a zero-current equilibrium condition at time
$0$, this relation can be inverted to obtain
$v(t)=\int_{0}^{t}dt'a(t-t')\phi(t')$, where $a(t)$ can be thought of
as the `admittance'. In the Laplace transform language,
$\tilde{\mu}(k)=\tilde{a}^{-1}(k)$, where $k$ is the Laplace variable
representing inverse time. Via the fluctuation-dissipation theorem,
they are related to the respective autocorrelation functions as
$\mu(t-t')=\langle\phi(t)\phi(t')\rangle_{v=0}$ and $a(t-t')=\langle
v(t)v(t')\rangle_{\phi=0}$.

The behaviour of $\mu(t)$ may be obtained by considering the polymer
segment on one side of the membrane only, say the right, with a sudden
introduction of $p$ extra monomers at the pore, corresponding to
impulse current $v(t)=p\delta(t)$. We then ask for the time-evolution
of the mean response $\langle\delta\Phi^{(r)}(t)\rangle$, where
$\delta\Phi^{(r)}(t)$ is the shift in chemical potential for the right
segment of the polymer at the pore. This means that for the
translocation problem (with both right and left segments), we would
have $\phi(t)=\delta\Phi^{(r)}(t)-\delta\Phi^{(l)}(t)$, where
$\delta\Phi^{(l)}(t)$ is the shift in chemical potential for the left
segment at the pore due to an opposite input current to it.
\begin{figure}[!t]
\includegraphics[width=0.7\linewidth,angle=270]{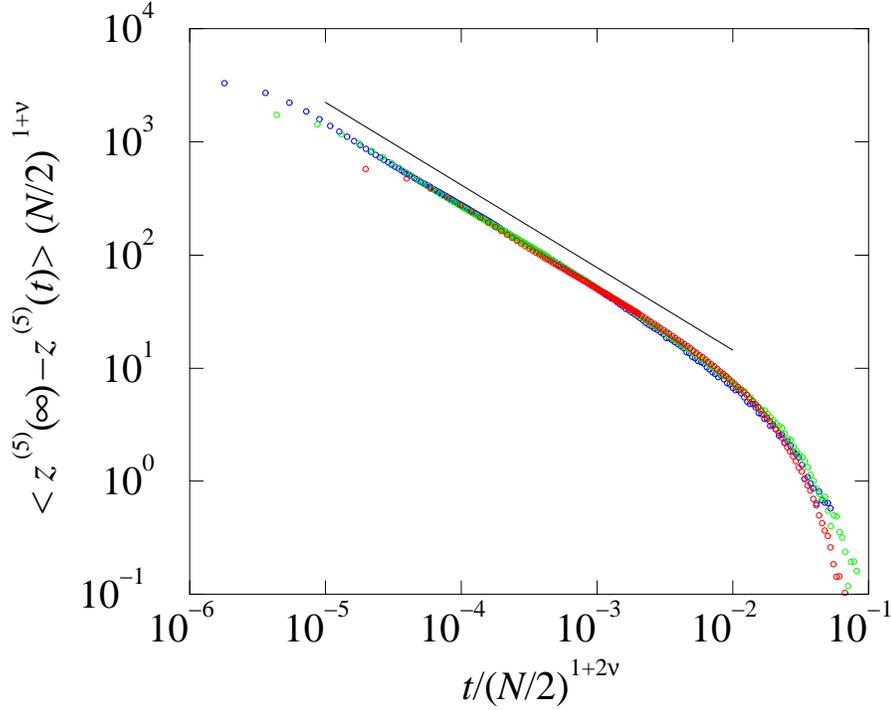}
\caption{Simulation results for the average chain tension component
perpendicular to the membrane proxied by $\langle
z^{(5)}(\infty)-z^{(5)}(t)\rangle$ following monomer injection at the
pore corresponding to $v(t)=p\delta(t)$, with $p=10$. See text for
details. Red circles: $N/2=50$, green circles: $N/2=100$, blue
circles: $N/2=150$, solid black line: $t^{-(1+\nu)/(1+2\nu)}$ with
$\nu=0.588$ for self-avoiding polymers. To obtain a data collapse, the
horizontal and vertical axes are scaled by $(N/2)^{1+2\nu}$ and
$(N/2)^{1+\nu}$, respectively.  The steeper drop at large times
correspond to the exponential decay $\exp(-t/\tau_{R})$.
\label{fig1}}
\end{figure}

We now argue that this mean response, and hence $\mu(t)$, takes the
form $\mu(t)\sim t^{-\alpha}\exp(-t/\tau_{R})$. The terminal
exponential decay $\exp(-t/\tau_{R})$ is expected from the relaxation
dynamics of the entire right segment of the polymer with one end
tethered at the pore \cite{longpaper}. To understand the physics
behind the exponent $\alpha$, we use the well-established result for
the relaxation time $t_{n}$ for $n$ self-avoiding Rouse monomers
scaling as $t_{n}\sim n^{1+2\nu}$. Based on the expression of $t_{n}$,
we anticipate that by time $t$ the extra monomers will be well
equilibrated across the inner part of the chain up to $n_{t}\sim
t^{1/(1+2\nu)}$ monomers from the pore, but not significantly
further. This internally equilibrated section of $n_{t}+p$ monomers
extends only $r(n_{t})\sim n_{t}^{\nu}$, less than its equilibrated
value $\left(n_{t}+p\right)^{\nu}$, because the larger scale
conformation has yet to adjust: the corresponding compressive force
from these $n_{t}+p$ monomers is expected by standard polymer scaling
\cite{degennes} to follow $f/(k_{B}T)\sim\delta
r(n_{t})/r^{2}(n_{t})\sim\nu p/\left[n_{t}r(n_{t})\right]\sim
t^{-(1+\nu)/(1+2\nu)}$, for $p \ll n_t$.  This force $f$ must be
transmitted to the  membrane, through a combination of decreased
tension at the pore and increased incidence of other membrane
contacts. The fraction borne by reducing chain tension at the pore
leads us to the inequality $\alpha\ge(1+\nu)/(1+2\nu)$, which is
significantly different from (but compatible with) the value
$\alpha_{1}=2/(1+2\nu)$ required to obtain $\tau_{d}\sim\tau_{R}$. It
seems unlikely that the adjustment at the membrane should be
disproportionately distributed between the chain tension at the pore
and other membrane contacts, leading to the expectation that the
inequality above is actually an equality.

We have confirmed this picture by measuring the impedance response
through simulations. In Ref. \cite{forced}, two of us have shown that
the centre-of-mass of the first few monomers is an excellent proxy for
chain tension at the pore and we assume here that this further
serves as a proxy for $\delta\Phi$. Based on this idea, we track
$\langle\delta\Phi^{(r)}(t)\rangle$ by measuring the distance of the average
centre-of-mass of the first $5$ monomers from the membrane, $\langle
z^{(5)}(t)\rangle$, in response to the injection of extra monomers
near the pore at time $0$. Specifically we consider the equilibrated
right segment of the polymer, of length $N/2-10$ (with one end
tethered at the pore), adding $10$ extra monomers at the tethered
end of the right segment at time $0$, corresponding to $p=10$,
bringing its length up to $N/2$. Using the proxy $\langle
z^{(5)}(t)\rangle$ we then track $\langle\delta\Phi^{(r)}(t)\rangle$. The
clear agreement between the exponent obtained from the simulation
results with the theoretical prediction of $\alpha=(1+\nu)/(1+2\nu)$
can be seen in Fig. \ref{fig1}. Note that the sharp
deviation of the data from the power law $t^{-(1+\nu)/(1+2\nu)}$ at
long times is due to the asymptotic exponential decay as
$\exp(-t/\tau_{R})$, as the data collapse shows.
\begin{figure}[!t]
 \begin{centering}
\includegraphics[angle=270,width=0.9\linewidth]{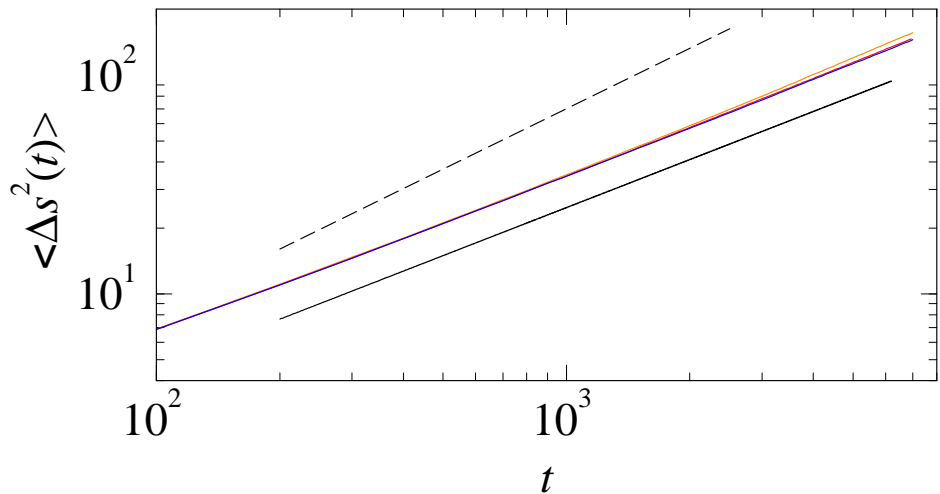} 
\par\end{centering}
\caption{Double-logarithmic plot of the mean squared
displacement of the reaction coordinate $\langle\Delta
s^{2}(t)\rangle$ as a function of time $t$, for $N=100$ (orange), 200
(red) and 500 (blue). The thick black line indicates the theoretically
expected slope corresponding to $\langle\Delta s^{2}(t)\rangle\sim
t^{(1+\nu)/(1+2\nu)}$. The dashed black line corresponds to
$\langle\Delta s^{2}(t)\rangle\sim t^{2/(1+2\nu)}$, which would have
been the slope of the $\langle\Delta s^2(t)\rangle$ vs. $t$ curve in
a double-logarithmic plot, if $\tau_d$ were to scale as $\tau_R\sim
N^{1+2\nu}$.
\label{fig2}}
\end{figure}

Having thus shown that $\mu(t)\sim
t^{-\frac{1+\nu}{1+2\nu}}\exp(-t/\tau_{R})$, we can expect that the
translocation dynamics is anomalous for $t<\tau_{R}$, in the sense
that the mean-square displacement of the monomers through the pore,
$\langle\Delta s^{2}(t)\rangle\sim t^{\beta}$ for some $\beta<1$ and
time $t<\tau_{R}$, whilst beyond the Rouse time it becomes simply
diffusive. The value $\beta=\alpha=\frac{1+\nu}{1+2\nu}$ follows
trivially by expressing $\langle\Delta s^{2}(t)\rangle$ in terms of
(translocative) velocity correlations $\left\langle
v(t)v(t')\right\rangle $, which (by the Fluctuation Dissipation
theorem) are given in terms of the time dependent admittance
$a(t-t')$, and hence inversely in terms of the corresponding
impedance.

Indeed, as shown in Fig. \ref{fig2}, a double-logarithmic plot of
$\langle\Delta s^{2}(t)\rangle$ vs. $t$ is consistent with
$\langle\Delta s^{2}(t)\rangle\sim t^{(1+\nu)/(1+2\nu)}$. The
behaviour of $\langle\Delta s^{2}(t)\rangle$ at short times is an
artifact of our model: at short times reptation moves dominate,
leading to a transport mechanism for {}``stored lengths''
\cite{rubinstein} along the polymer's contour in which individual
units of stored length cannot pass each other. As a result, the
dynamics of $s(t)$, governed by the movement of stored length units
across the pore, is equivalent to a process known as {}``single-file
diffusion'' on a line, characterized by the scaling $\langle\Delta
s^{2}(t)\rangle\sim t^{1/2}$ (not shown here). At long times the
polymer tails will relax, leading to $\langle\Delta
s^{2}(t)\rangle\sim t$ for $t>\tau_{R}$. The presence of two
crossovers, the first one from $\langle\Delta s^{2}(t)\rangle\sim
t^{1/2}$ to $\langle\Delta s^{2}(t)\rangle\sim t^{(1+\nu)/(1+2\nu)}$
and the second one from $\langle\Delta s^{2}(t)\rangle\sim
t^{(1+\nu)/(1+2\nu)}$ to $\langle\Delta s^{2}(t)\rangle\sim t$ at
$t\approx\tau_{R}$, complicates the precise numerical verification of
the exponent $(1+\nu)/(1+2\nu)$. However, as shown in Fig. \ref{fig2},
there is an extended regime in time at which the quantity
$t^{-(1+\nu)/(1+2\nu)}\langle\Delta s^{2}(t)\rangle$ is nearly
constant.

The subdiffusive behaviour $\langle\Delta s^{2}(t)\rangle\sim
t^{\frac{1+\nu}{1+2\nu}}$ for $t<\tau_{R}$, combined with the
diffusive behaviour for $t\geq\tau_{R}$ leads to the dwell time
scaling as $\tau_{d}\sim N^{2+\nu}$, based on the criterion that
$\sqrt{\langle\Delta s^{2}(\tau_{d})\rangle}\sim N$.  The dwell time
exponent $2+\nu\simeq2.59$ is in acceptable agreement with the two
numerical results on $\tau_{d}$ in 3D as mentioned in the introduction
of this paper, and in Table I below we present new high-precision
simulation data in support of $\tau_{d}\sim N^{2+\nu}$, in terms of
the median unthreading time. The unthreading time $\tau_u$ is defined
as the time for the polymer to leave the pore with $s(t=0)=N/2$ and
the two polymer segments equilibrated at $t=0$. Both $\tau_u$ and
$\tau_d$ scale the same way, since $\tau_u<\tau_d<2\tau_u$
\cite{longpaper}.
\begin{center}
\begin{tabular}{p{2cm}|p{2cm}|p{2cm}}
\hspace{8mm}$N$ &
\hspace{8mm}$\tau_u$ &
\hspace{4mm}$\tau_u/N^{2+\nu}$ \tabularnewline
\hline
\hline 
\hspace{7mm}100 &
\hspace{6mm}65136 &
\hspace{6mm}0.434 \tabularnewline
\hline 
\hspace{7mm}150 &
\hspace{5mm}183423 &
\hspace{6mm}0.428 \tabularnewline
\hline 
\hspace{7mm}200 &
\hspace{5mm}393245 &
\hspace{6mm}0.436 \tabularnewline
\hline 
\hspace{7mm}250 &
\hspace{5mm}714619 &
\hspace{6mm}0.445 \tabularnewline
\hline 
\hspace{7mm}300 &
\hspace{4mm}1133948&
\hspace{6mm}0.440\tabularnewline
\hline 
\hspace{7mm}400 &
\hspace{4mm}2369379&
\hspace{6mm}0.437\tabularnewline
\hline 
\hspace{7mm}500 &
\hspace{4mm}4160669&
\hspace{6mm}0.431\tabularnewline
\hline
\end{tabular}

\vspace{3mm}
{\footnotesize Table I: Median unthreading time over 1,024 runs for
each $N$.} \vspace{3mm}
\par\end{center}

We now reflect on the theory presented in Ref. \cite{dubbeldam}.

We have defined $\tau_{d}$ as the pore-blockade time in experiments;
i.e., if we define a state of the polymer with $s(t)=0$ as `0'
(polymer just detached from the pore on one side), and with $s(t)=N$
as `N', then $\tau_{d}$ is the first passage time required to travel
from state 0 to state N \textit{without\/} possible reoccurances of
state 0. In Ref. \cite{dubbeldam}, the authors attach a bead at the
$s=0$ end of the polymer, preventing it from leaving the pore,
creating a situation where the polymer returns to state 0 multiple
number of times before it eventually reaches state N. The repeated
returns to state 0 implies that by construction of the problem, the
polymer encounters a free energy barrier on its way from state 0 to
$s=N/2$, where the polymer's configurational entropy is the
lowest. The authors then proceed to express their translocation time
($\tau_{t}$ hereafter), defined as the first passage time required to
travel from state 0 to state N \textit{with\/} reoccurances of state 0,
in terms of this free energy barrier. Below we settle the differences
between $\tau_{t}$ of Ref. \cite{dubbeldam} and our $\tau_{d}$.

Consider the case where we attach a bead at $s=0$ and another at
$s=N$, preventing it from leaving the pore. We then characterize the
state of the polymer as follows.  At state x and x$'$ the
polymer can have all values of $s$ except $0,N/2$ and $N$; and at
states m and m$'$, $s=N/2$. The notational distinction between primed
and unprimed states is that a primed state can occur only between two
consecutive states 0, or between two consecutive states N, while an
unprimed state occurs only between state 0 and state N.
Its dynamics is then given by the sequence of states, e.g., 
\begin{eqnarray}
...\mbox{N \!x\! m\! x}\overbrace{\mbox{0 \!x$'$0
      \!x$'$m$'$x$'$m$'$x$'$0 \!x$'$}\underbrace{\mbox{0 \!x \!m \!x
	\!m \!x \!m \!x
	\!N}}_{\tau_{d}}}^{\tau_{t}}\mbox{x$'$N}...
\nonumber
\end{eqnarray}
where the corresponding times taken ($\tau_{t}$ and $\tau_{d}$) are
indicated. Note in the above definitions that $\tau_t>\tau_d$: since,
due to the presence of the entropic barrier as described above,
$\tau_{t}$ includes the extra time spent in between the first and the
last occurrence of state 0 before the polymer eventually proceeds to
state N. In other words, {\it $\tau_t$ includes the effect of the
entropic barrier, while $\tau_d$ does not}.
A probability argument then leads us to
\begin{eqnarray}
\frac{\tau_{t}}{\tau_{d}}\,=\,\frac{1}{p_{\text{x}}+p_{\text{m}}}\,=\,\frac{f_{\text{x}}\,(1+f_{\text{m}})}{(p_{\text{m}}+p_{\text{m$'$}})f_{\text{m}}(1+f_{\text{x}})}\,,
\label{e1}
\end{eqnarray}
where $p_{\text{m}}$, $p_{\text{m}'}$ and $p_{\text{x}}$ are the
probabilities of the corresponding states,
$f_{\text{m}}=p_{\text{m}}/p_{\text{m}'}$ and
$f_{\text{x}}=p_{\text{m}}/p_{\text{x}}$. The partition sum of a
polymer of length $n$ with one end tethered on a membrane is given by
$Z_{n}\sim\lambda^{n}~n^{\gamma_{1}-1}$ with $\lambda$ a non-universal
constant and $\gamma_{1}=0.68$ \cite{diehla}, and therefore we have
$p_{\text{m}}+p_{\text{m}'}=Z_{N/2}^{2}/\left[\sum_{s=0}^{N}Z_{s}Z_{N-s}\right]\sim1/N$.
Similarly, $f_{\text{x}}\sim1/N$ \cite{wolt}. Finally,
$f_{\text{m}}\approx1$ \cite{note} yields $\tau_{t}\sim\tau_{d}$.

In Ref. \cite{dubbeldam} the authors include a factor $N^{1-\gamma_1}$
in $\tau_t$ to account for the effect of the entropic
barrier. However, we have shown above that $\tau_{t}\sim\tau_{d}$,
i.e., the free energy barrier does not play a role for the scaling
behaviour of $\tau_{t}$ with $N$. This implies, since $\tau_t$
includes the effect of the entropic barrier and $\tau_d$ does not,
that the theoretical expression for $\tau_{t}$ in
Ref. \cite{dubbeldam} cannot be correct. The numerical result
$\tau_{t}\sim N^{2.52\pm0.04}$ in Ref. \cite{dubbeldam}, however,
confirms our theoretical expression $\tau_{d}\sim N^{2+\nu}$.

To conclude, in this paper we have characterized the anomalous
dynamics of unbiased translocation and obtained the scaling of the
dwell time in terms of the polymer length. In future work, we will
study the role of hydrodynamics. Rouse friction may be an appropriate
model for the dynamics of long biopolymers in the environment within
living cells, if it is sufficiently gel-like to support screened
hydrodynamics on the timescale of their configurational
relaxation. However, we should also ask what is expected in the other
extreme of unscreened (Zimm) hydrodynamics. For our theoretical
discussion the key difference is that, instead of the Rouse time
$\tau_{R}$, in the Zimm case the configurational relaxation times
scale with $N$ according to $\tau_{\text{Zimm}}\sim N^{3\nu}$ in 3D,
which upon substitution into our earlier argument would gives the
lower bound value $\alpha=(1+\nu)/(3\nu)$ for the time exponent of the
impedance, leading to $\tau_{d}\sim N^{1+2\nu}$ (whose resemblance to
the Rouse time is a coincidence --- note that with hydrodynamics Rouse
time loses all relevance). These results, however, do need to be
verified by simulations incorporating hydrodynamics.


\begin{thebibliography}{10}
\bibitem{note} $f_{\text{m}}$ is a number slightly smaller than
$1$: if the polymer reaches $s=N/2$ from state 0, due to the memory
effects, it will have slightly higher chance to go back to state 0
rather than to proceed to state N \cite{wolt}.
\end{thebibliography}
\end{document}